\RequirePackage{lineno}
\documentclass[review]{elsarticle}

\modulolinenumbers[5]

\usepackage{graphicx}
\usepackage{dcolumn}
\usepackage{bm}
\usepackage{xfrac}
\usepackage{geometry}

\usepackage[printwatermark]{xwatermark}
\usepackage{xcolor}

\usepackage{graphicx,amsmath,amssymb}
\usepackage{epstopdf}
\usepackage{lineno}
\usepackage[colorlinks=true, urlcolor=blue, citecolor=blue]{hyperref}

\begin{document}


\title{Measurement of the $^3$He Spin-Structure Functions and of Neutron $(^3$He$)$ Spin-Dependent Sum Rules at 0.035~$\le$~$Q^{2}$~$\le$~0.24~GeV$^{2}$}

\author{Jefferson Lab E97-110 Collaboration}
\address{}
\author{V.~Sulkosky}
\address{College of William and Mary, Williamsburg, Virginia 23187-8795, USA }
\address{Thomas Jefferson National Accelerator Facility, Newport News, Virginia 23606, USA }
\address{University of Virginia, Charlottesville, Virginia 22904, USA}
\author{J.~T.~Singh}
\address{University of Virginia, Charlottesville, Virginia 22904, USA}
\author{C.~Peng}
\address{Duke University, Durham, North Carolina 27708, USA}
\author{J.-P.~Chen}
\address{Thomas Jefferson National Accelerator Facility, Newport News, Virginia 23606, USA }
\author{A.~Deur\footnote{Contact author. Email: deurpam@jlab.org}}
\address{University of Virginia, Charlottesville, Virginia 22904, USA}
\address{Thomas Jefferson National Accelerator Facility, Newport News, Virginia 23606, USA }
\author{S.~Abrahamyan}
\address{Yerevan Physics Institute, Yerevan 375036, Armenia}
\author{K.~A.~Aniol}
\address{California State University, Los Angeles, Los Angeles, California 90032, USA}
\author{D.~S.~Armstrong}
\address{College of William and Mary, Williamsburg, Virginia 23187-8795, USA }
\author{T.~Averett}
\address{College of William and Mary, Williamsburg, Virginia 23187-8795, USA }
% ...


%
\begin{abstract}
The spin-structure functions $g_1$  and $g_2$, and 
the spin-dependent partial cross-section $\sigma_\mathrm{TT}$ 
have been extracted from  the polarized cross-sections differences,
$\Delta \sigma_{\parallel}\hspace{-0.06cm}\left(\nu,Q^{2}\right)$ and 
$\Delta \sigma_{\perp}\hspace{-0.06cm}\left(\nu,Q^{2}\right)$ measured for the 
$\vec{^\textrm{3}\textrm{He}}(\vec{\textrm{e}},\textrm{e}')\textrm{X}$
reaction, in the E97-110 experiment at Jefferson Lab.  
Polarized electrons with energies from 1.147 to 4.404 GeV were scattered at 
angles of 6$^{\circ}$ and 9$^{\circ}$ from a longitudinally or transversely 
polarized $^{3}$He target.
The data cover the kinematic regions of the quasi-elastic, resonance production and beyond.  From the extracted 
spin-structure functions, the first moments $\overline{\Gamma_1}\hspace{-0.06cm}\left(Q^{2}\right)$, 
 $\Gamma_2\hspace{-0.06cm}\left(Q^{2}\right)$ and $I_{\mathrm{TT}}\hspace{-0.06cm}\left(Q^{2}\right)$ 
 are evaluated with high precision for the neutron 
 in the $Q^2$ range from 0.035 to 0.24~GeV$^{2}$.
 The comparison of the data and the chiral effective field theory predictions 
reveals the importance of proper treatment of the $\Delta$ degree of freedom for spin observables.
\end{abstract}

\maketitle
The study of nucleon spin structure has been actively pursued over the past
thirty years~\cite{Chen:2005tda}, both 
theoretically and experimentally at several laboratories, including 
CERN~\cite{Ashman:1987hv}, 
SLAC~\cite{slac, Anthony:2002hy}, 
DESY~\cite{Ackerstaff:1997ws, Airapetian:2002wd} and 
Jefferson Lab (JLab)~\cite{Yun:2002td,Slifer:2008re,Amarian:2002ar,Amarian:2003jy,Amarian:2004yf,Deur:2004ti,Slifer:2008xu,Solvignon:2013yun,Adhikari:2017wox} 
using doubly polarized inclusive lepton scattering. This research
provides a powerful means to study the strong force and its gauge theory, quantum chromodynamics (QCD).
They are well tested at high momenta where perturbative expansions in $\alpha_s$, QCD's coupling, are feasible. 
Extensive data also exist at intermediate momenta. Yet, at the low momenta characterizing the domain of quark
confinement,  
there are no precision data.  
There, studies are complicated by 1) the difficulty of finding calculable observables, and 2)  the inapplicability
of perturbative QCD due to the steep increase of $\alpha_s$~\cite{Deur:2016tte}.
Sum rules offer a remarkable opportunity to address the first problem by equating measurable moments of structure functions
to calculable Compton scattering amplitudes.
The second challenge demands the use of non-perturbative techniques such as lattice QCD, 
or of effective approaches such as chiral effective field theory ($\chi$EFT)~\cite{Bernard:1995dp}.
In  $\chi$EFT, the effective hadronic  degrees of freedom, relevant at low momenta, are used
--rather than the fundamental ones (partons) explicit only at large momenta-- and the $\chi$EFT Lagrangian structure is
established by the symmetries of QCD.

A spin-dependent sum rule of great interest is the one of Gerasimov, Drell, and Hearn
(GDH)~\cite{Gerasimov:1965et}.  It links an
integral over the excitation spectrum of the helicity-dependent
photoabsorption cross-sections to the target's anomalous magnetic
moment $\kappa$.  
The sum rule stems from causality, unitarity, and
Lorentz and gauge invariances. Its expression for a spin-$\sfrac{1}{2}$ target is:
\begin{equation}
{\int_{\nu_{0}}^{\infty} \left[\sigma_{\sfrac{1}{2}}(\nu) - \sigma_{\sfrac{3}{2}}(\nu)\right] \frac{d\nu}{\nu} =  - \frac{2\pi^{2}\alpha}{M_t^{2}}\kappa^{2}},
\label{eq:gdh}
\end{equation}
where ${M_t}$ is the target mass, $\nu$ the photon
energy, $\nu_{0}$ the inelastic threshold and $\alpha$ is the fine-structure constant.  The
$\sfrac{1}{2}~(\sfrac{3}{2})$ indicates that the photon
helicity is parallel (anti-parallel) to the target spin.  
The GDH sum rule can be applied to various polarized targets such as $^3$He and the neutron, with predictions of -498.0 and -232.5 $\mu$b, respectively.
The sum rule was verified on the proton by the MAMI, ELSA, and LEGS experiments~\cite{Ahrens:2001qt}
with circularly polarized photons of up to $\nu\approx3$ GeV.

Starting in the 1980's, generalizations of the integrand for virtual photon absorption were
proposed~\cite{Anselmino:1988hn,Drechsel:2000ct,Drechsel:2002ar}, e.g.:
%
\begin{align}
\hspace{-0.1cm} I_\mathrm{TT}(Q^2)\hspace{-0.06cm} & \equiv \hspace{-0.1cm} \frac{M_t^2}{8\pi^2\alpha} \hspace{-0.1cm} \int_{\nu_0}^{\infty} \hspace{-0.1cm}\frac{\kappa_f(\nu,Q^2)}{\nu} \hspace{-0.04cm} \frac{\sigma_{1/2}(\nu,Q^2) \hspace{-0.06cm} -  \hspace{-0.06cm} \sigma_{3/2}(\nu,Q^2)}{\nu}d\nu \nonumber \\
  & \hspace{-0.2cm} =\frac{2M_t^2}{Q^2}\hspace{-0.15cm}\int_0^{x_0}\hspace{-0.08cm}\Bigr[g_1(x,Q^2)\hspace{-0.08cm}-\hspace{-0.08cm}\frac{4M_t^2}{Q^2}x^2g_2(x,Q^2)\hspace{-0.04cm}\Bigl]dx, \label{eq:gdhsum_def1}
\end{align}
where  $\nu$ is the energy transfer, $Q^{2}$ the four-momentum transfer squared, 
$x=\sfrac{Q^{2}}{2M_t\nu}$ is the Bjorken scaling variable, $x_{0} =\sfrac{Q^{2}}{2M_t\nu_{0}}$, and $g_1$ and $g_2$ are the spin structure functions.
 $\kappa_f$, the virtual photon flux, normalizes the partial cross-sections $\sigma_{\sfrac{1}{2},\sfrac{3}{2}}$~\cite{Chen:2005tda}.
Its form is conventional and we will use here the Hand convention~\cite{Hand:1963bb}, $\kappa_f = \nu-\sfrac{Q^2}{2M}$.
Different choices of convention have lead to different generalization of the GDH sum~\cite{Drechsel:2000ct}.
However, the value of $I_\mathrm{TT}(Q^2)$ is independent of the choice of $\kappa_f$ since 
it also normalizes the  $\sigma_{\sfrac{1}{2},\sfrac{3}{2}}$, as shown explicitly when $I_\mathrm{TT}(Q^2)$ is
expressed with $g_1$ and $g_2$.
These relations extend the integrand to $Q^2>0$.
The sum rule itself was generalized by Ji and Osborne~\cite{Ji:1999mr} using a dispersion relation involving 
the forward virtual Compton scattering amplitude $S_1(\nu,Q^{2})$ in the $\nu \to 0$ limit:
%
\begin{equation}
\overline{\Gamma_1}\left(Q^{2}\right) \equiv \int_{0}^{x_{0}} g_1(x,Q^{2})dx = \frac{Q^{2}}{8}\overline{S_1}(0,Q^{2})\,,
\label{eq:gengdh}
\end{equation}
where the bar indicates exclusion of the elastic contribution.  This relation, valid
at any $Q^{2}$, can be applied back to Eq.~(\ref{eq:gdhsum_def1}), equating the moment $I_{TT}(Q^2)$ to $A_{TT}(\nu,Q^2)$,
the spin-flip   doubly virtual Compton scattering amplitude in the $\nu \to 0$ limit.
 The amplitudes $\overline{S_1}(0,Q^{2})$ and $A_{TT}(0, Q^2)$ are calculable, e.g. in QCD as four-point functions 
using lattice techniques~\cite{Chambers:2017dov}, or by $\chi$EFT.
Eqs.~(\ref{eq:gdhsum_def1}) or (\ref{eq:gengdh}) can then be used to compare these calculations to experimental data. 
Such data became available at intermediate~\cite{Yun:2002td, Slifer:2008re, Amarian:2002ar, Amarian:2003jy, Amarian:2004yf,Deur:2004ti} 
and large $Q^2$~\cite{Airapetian:2002wd} in the 1990s and 2000s.
Their lowest $Q^2$ points revealed tensions with the available
$\chi$EFT calculations of $\overline{S_1}(0,Q^{2})$ and $A_{TT}(0, Q^2)$~\cite{Bernard:2002pw,Ji:1999pd}.
The discrepancies between data and calculations can be due to the $Q^2$ coverage of the experiments being 
not low enough for a valid comparison with $\chi$EFT, and/or to the calculations themselves. 
 The data~\cite{Amarian:2002ar,Amarian:2003jy,Amarian:2004yf} showed the importance for $\chi$EFT calculations
to account for the first excited state (the $\Delta (1232)$) beyond the nucleon ground state.
The data also revealed the need for measuring spin moments at $Q^2$ low enough so that $\chi$EFT calculations
can be accurately tested. 

The other spin structure function $g_2$  is expected to obey the 
Burkhardt--Cottingham (BC) sum rule~\cite{Burkhardt:1970ti}: 
\begin{equation}
\label{BCSUMRULE}
\Gamma_2(Q^2) \equiv \int_0^1 g_2(x,Q^2) dx = 0\,,
\end{equation}
a super-convergence relation, i.e. implicitly independent of $Q^2$,
derived from the dispersion relation for the Compton scattering amplitude
$S_2\left(Q^{2}\right)$~\cite{Drechsel:2002ar}.  The BC sum
rule's validity depends on the convergence of the integral and assumes that
$g_2$ is well-behaved as $x\to 0$~\cite{Jaffe:1990qh}. 

\begin{figure}
  \centering
    \includegraphics[width=0.50\textwidth]{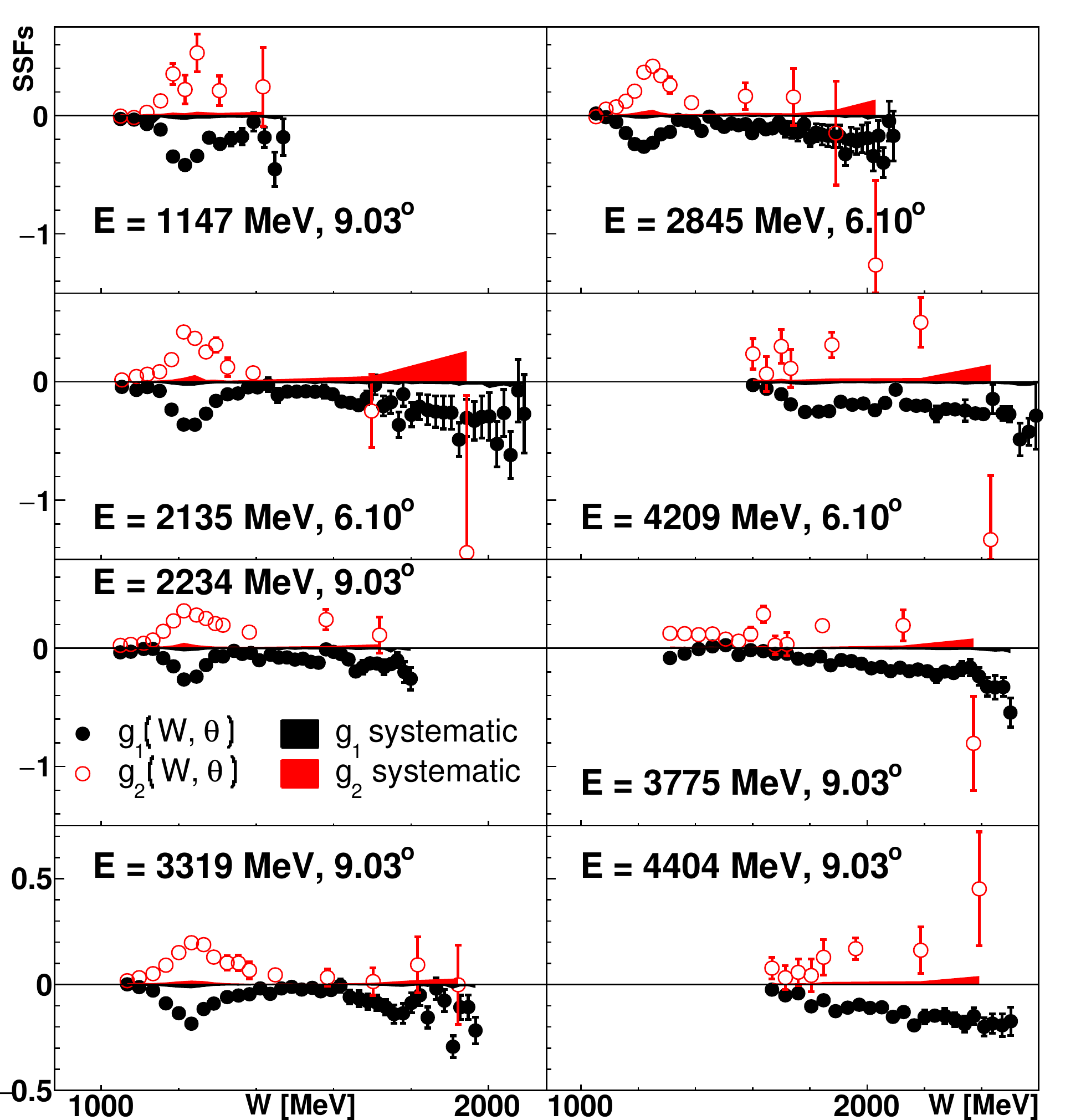} 
  \caption{\label{3Hessf} Spin structure functions (SSFs) $g_1^{^3\mbox{He}}$  and $g_2^{^3\mbox{He}}$ at fixed $\theta$ and $E$, versus $W$.
    The error bars (bands) provide the statistical (systematic) uncertainty. }
\end{figure}

We present here data on $g_1$, $g_2$ and
$\sigma_\mathrm{TT} \equiv (\sigma_{\sfrac{1}{2}} - \sigma_{\sfrac{3}{2}})/2$ 
on $^{3}$He, 
and of  $\overline{\Gamma_1}$,  $\Gamma_2$ and $I_\mathrm{TT}$ for the neutron, for
$0.035 \le Q^{2} \le$ 0.24~GeV$^{2}$ 
 from experiment E97-110~\cite{E97110,VINCETHESIS}. Data were acquired in Hall
A~\cite{Alcorn:2004sb} at JLab.  
We measured the inclusive reaction $\vec{\rm{^{3}He}}$($\vec{\rm{e}},\rm{e'}$) with a longitudinally
polarized electron beam scattered from longitudinally or transversely (in-plane) 
polarized $^{3}$He~\cite{Alcorn:2004sb}.  Eight beam energies $E$ and two scattering angles 
$\theta$ were used to cover kinematics at constant $Q^{2}$, see Fig.~\ref{3Hessf}.
%
The data cover invariant mass $W = \sqrt{M^2 + 2M\nu - Q^2}$ ($M$ is the nucleon mass) 
values from the elastic 
up to 2.5~GeV;  however, only the results above the pion production
threshold ($W = 1.073$~GeV) are discussed here.  
Spin asymmetries and absolute cross-sections were both  measured.   
The beam polarization was flipped pseudo-randomly at 30 Hz and M\o ller and Compton polarimeters~\cite{Alcorn:2004sb} 
measured it to average at 75.0~$\pm$~2.3\%.  
The  beam current ranged from 1 to 10~$\mu$A depending on the trigger rate.  
The data acquisition rate was limited to 4~kHz to keep the deadtime below 20\%.   

The $^{3}$He target was polarized by spin-exchange optical pumping 
(SEOP)~\cite{Gentile:2016uud}. 
Two sets of Helmholtz coils providing a parallel or transverse 2.5~mT uniform 
field allowed us to orient the $^{3}$He spins longitudinally or perpendicularly 
to the beam direction. The target had about 12~atm of $^3$He gas 
in a glass cell consisting of two connected chambers.  
The SEOP process occurred in the upper chamber, which was illuminated with 90~W of 
laser light at a wavelength of 795~nm.  The electron beam passed through a lower chamber made of a
40~cm-long cylinder with a diameter of 2~cm and hemispherical glass windows 
at both ends.
Two independent polarimetries monitored the $^3$He polarization: 
nuclear magnetic resonance (NMR) 
and electron paramagnetic resonance (EPR). 
The NMR system was calibrated using adiabatic fast passage and the known 
thermal equilibrium polarization of water. The 
polarization was independently cross-checked by measuring the elastic $^3$He 
asymmetry. The average in-beam target polarization was (39.0~$\pm$~1.6)\%.

The scattered electrons were detected by a High
Resolution Spectrometer (HRS)~\cite{Alcorn:2004sb} with a lowest 
scattering angle reachable of 12.5$^{\circ}$. A horizontally-bending dipole magnet ~\cite{septum} 
was placed in front of the HRS so that electrons with scattering angles of
6$^{\circ}$ or 9$^{\circ}$ could be detected. The HRS detector package consisted of
a pair of drift chambers for tracking, a pair of scintillator planes for 
triggering and a gas Cherenkov counter, together with a two
layer electromagnetic calorimeter for particle identification.  
Details of the experimental set-up and its performance can be found in~\cite{E97110,VINCETHESIS}.

The $g_1$ and $g_2$ spin structure functions were extracted 
from the cross-section differences $\Delta\sigma_{\parallel} \equiv
\frac{d^2\sigma^{\downarrow\Uparrow}}{d\Omega dE'} -
\frac{d^2\sigma^{\uparrow\Uparrow}}{d\Omega dE'}$ and 
$\Delta\sigma_{\perp} \equiv \frac{d^2\sigma^{\downarrow\Rightarrow}}{d\Omega
  dE'} - \frac{d^2\sigma^{\uparrow\Rightarrow}}{d\Omega dE'}$ for the case
where the target polarization is aligned parallel or perpendicular, respectively,
to the beam direction:
\begin{align}
\nonumber
g_1 & = \frac{MQ^2\nu}{4\alpha^2}\frac{E}{E'}\frac{1}
{E + E'}\left[\Delta\sigma_{\parallel} + \tan\left(\frac{\theta}{2}\right)\Delta\sigma_{\perp} \right]\,
\\
g_2 & = \hspace{-0.05cm}\frac{MQ^2\nu}{8\alpha^2E'(E + E')}\left[-\Delta\sigma_{\parallel} + \frac{E + E'\cos~\theta}{E'\sin~\theta}\Delta\sigma_{\perp} \right].
\nonumber
\end{align}
The cross-section differences $\Delta \sigma_{\parallel,\perp}$ were formed by combining 
longitudinal and transverse asymmetries $A_{\parallel}$ and $A_{\perp}$ with the
 unpolarized absolute cross-section $\sigma_{0}$:  \mbox{$\Delta
\sigma_{\parallel,\perp}$ = 2$\sigma_{0}A_{\parallel,\perp}$}.  
Unpolarized backgrounds cancel in $\Delta\sigma$
 and polarized background are negligible since only $^3$He nuclei are significantly polarized.
The asymmetries were corrected for the beam and target polarizations,
as well as beam charge and data acquisition lifetime asymmetries.
The dilution of the asymmetry by unpolarized background canceling that 
 same background in  $\sigma_0$, such correction is unnecessary when forming $\Delta \sigma$

The absolute cross-section was obtained by correcting for the
finite HRS acceptance and detector inefficiencies.
The $1/\nu$ weighting of the GDH sum emphasizes low $\nu$ contributions. Thus,
contamination from elastic and quasi-elastic events appearing beyond the electroproduction
threshold due to detector resolution and radiative tails was carefully studied and corrected on both $\sigma_{0}$
and $\Delta \sigma_{\parallel,\perp}$.
The high HRS momentum resolution helped to minimize the contamination. 
For the neutron moments, the quasi-elastic contamination was studied and subtracted by building a model of our 
data with guidance from state-of-the-art Faddeev calculations~\cite{Golak:2002fk} and the MAID~\cite{Drechsel:1998hk} model. 
The estimated uncertainty from the subtraction and the effect of varying the lower limit of integration 
(to account for below-threshold pion production) were included in our systematic uncertainty.
Since $g_1$ and $g_2$ are defined in the Born approximation, 
radiative corrections were applied following Ref.~\cite{Mo:1968cg} for the unpolarized case and 
using Ref.~\cite{Akushevich:1994dn} to
include polarized effects.  In the unfolding procedure described in [36], cross-section model or data at lower energy
are required. To avoid a model-dependent systematic uncertainty, lower energy data gathered for that purpose during the experiment
were used in the unfolding procedure.

\begin{figure}
  \centering
\includegraphics[width=0.5\textwidth]{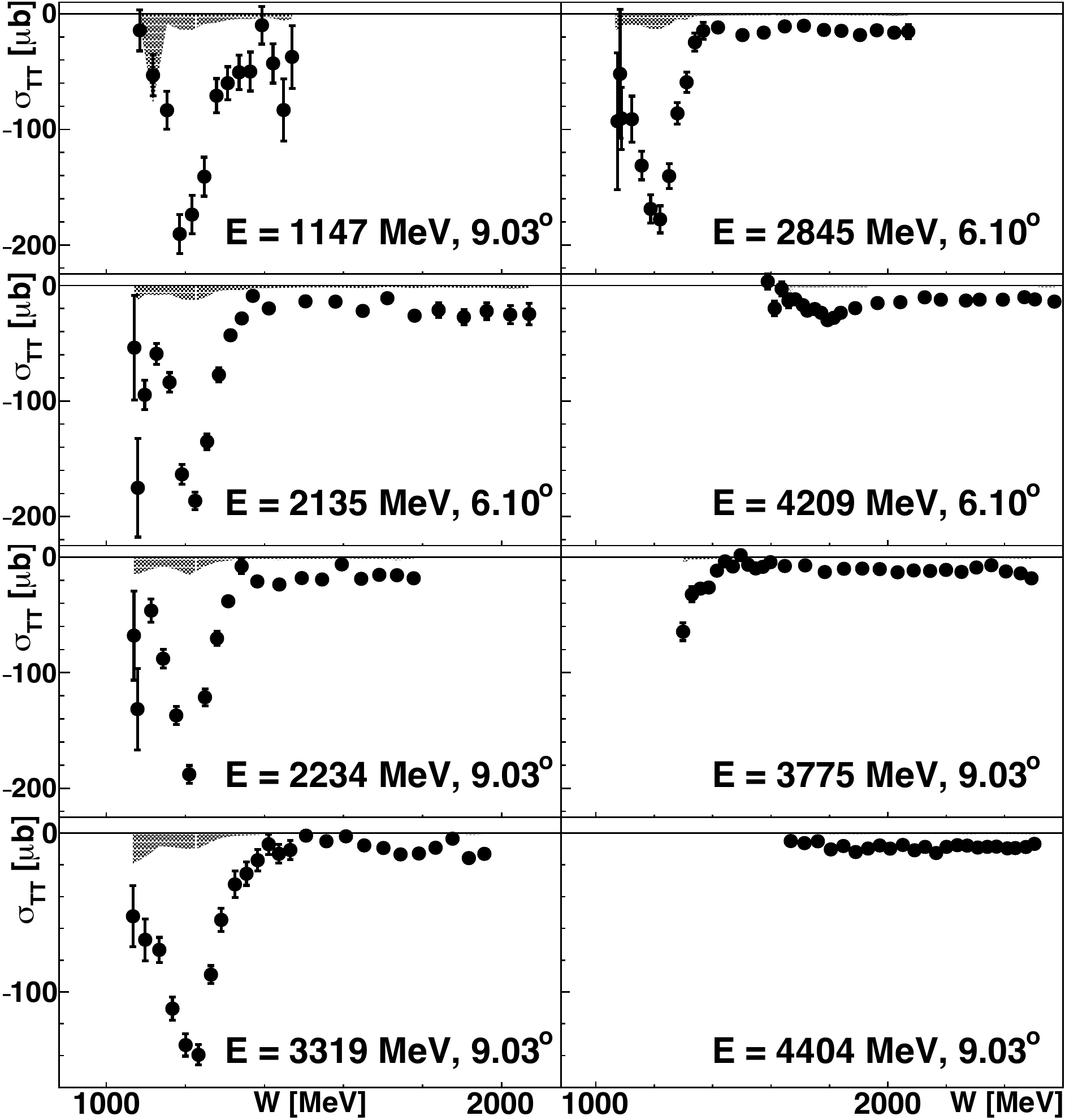} 
  \caption{\label{3Hestt} $\sigma_\mathrm{TT}^{^3\mbox{He}}$
  at fixed $\theta$ and $E$, versus $W$.
  The error bars (bands) provide the statistical (systematic) uncertainty. 
  }  
\end{figure}

The results for $g_1$  and $g_2$, and for $\sigma_\mathrm{TT}$
on $^{3}$He are shown in Fig.~\ref{3Hessf} and 
Fig.~\ref{3Hestt}, respectively.
The data are provided from the pion threshold. The error bars represent the statistical uncertainty. 
Systematic uncertainties are shown by the lower band for $g_1$ and $\sigma_\mathrm{TT}$ or the upper band for $g_2$.
The main systematic uncertainties are from the absolute
cross-sections (3.5 to 4.5\%), beam polarization (3.5\%), target polarization (3 to 5\%) and
radiative corrections (3 to 7\%). 
When combining uncertainties, the uncorrelated ones are added in quadrature. 
The correlated ones are added linearly. 
The full systematic uncertainty, shown by the band in Figs.~\ref{3Hessf} and~\ref{3Hestt}, 
is the uncorrelated and correlated uncertainties added quadratically.
The total systematic for $g_1$ varies between 12\% at low $W$ to
9\% at high $W$, for $g_2$ it is about 13\% over the whole $W$ range, and
for $\sigma_\mathrm{TT}$ between 11\% at low $W$ to 8\% at high $W$.
The data display a prominent feature in the $\Delta (1232)$ region. 
There, $g_1 \approx - g_2$.  This is expected, since the $\Delta$ is an $M1$ resonance for which 
the longitudinal-transverse interference cross-section
$\sigma'_\mathrm{LT} \propto \left(g_1 + g_2 \right)$ is anticipated to be highly 
suppressed~\cite{Drechsel:2000ct}. Above the $\Delta$, both spin structure functions 
decrease in magnitude, to increase again as $W$ approaches 2~GeV while still
displaying an approximate symmetry indicating the smallness of $\sigma'_\mathrm{LT}$.

To obtain $\overline{\Gamma_1^n}$ and $I_\mathrm{TT}^n$, we evaluated $g_1$, $g_2$ and $\sigma_\mathrm{TT}$
 at constant $Q^{2}$ by interpolating 
the fixed $\theta$ and $E$ data.
The moments  were then formed for each value of $Q^{2}$ with integration limits
from pion threshold to  the lowest $x$ value experimentally covered, see tables of the Supplemental Material.
The same neutron parameterization as used in Ref.~\cite{Adhikari:2017wox} was used
to complete the integration down to $x = 0.001$, and the recent Regge parameterization~\cite{Bass:2018uon} 
was used for $x < 0.001$. The unmeasured part is about 10\% of the full moments.
The parameters of the extrapolation models were varied within their estimated ranges, and the variations were combined 
into the extrapolation uncertainty. 

The neutron moments were obtained using the prescription in 
Ref.~\cite{CiofidegliAtti:1996cg} which treats the polarized $^{3}$He nucleus as an effective polarized neutron.  
 The resulting uncertainty is 6 to 14\%, the higher
uncertainties corresponding to our lowest $Q^{2}$ values.
Results for the integrals are given in  the tables 
of the Supplemental Material.

In Fig.~\ref{moments} our $\overline{\Gamma_1^{n}}$ is compared to $\chi$EFT calculations~\cite{Ji:1999pd, Bernard:2012hb,Lensky:2014dda}, 
models~\cite{Burkert:1992tg,Pasechnik:2010fg}, the MAID phenomenological parameterization~\cite{Drechsel:1998hk}
which contains only  resonance contributions, and
earlier data~\cite{Yun:2002td,Amarian:2003jy}.  
Where the $Q^2$ coverages overlap, our data agree with the earlier data extracted either from the deuteron or $^3$He. 
Our precision is much improved compared to the EG1 data~\cite{Yun:2002td} and similar to that of the E94-010~\cite{Amarian:2003jy} data at larger $Q^2$.  
\begin{figure}
  \centering
    \includegraphics[width=0.5\textwidth]{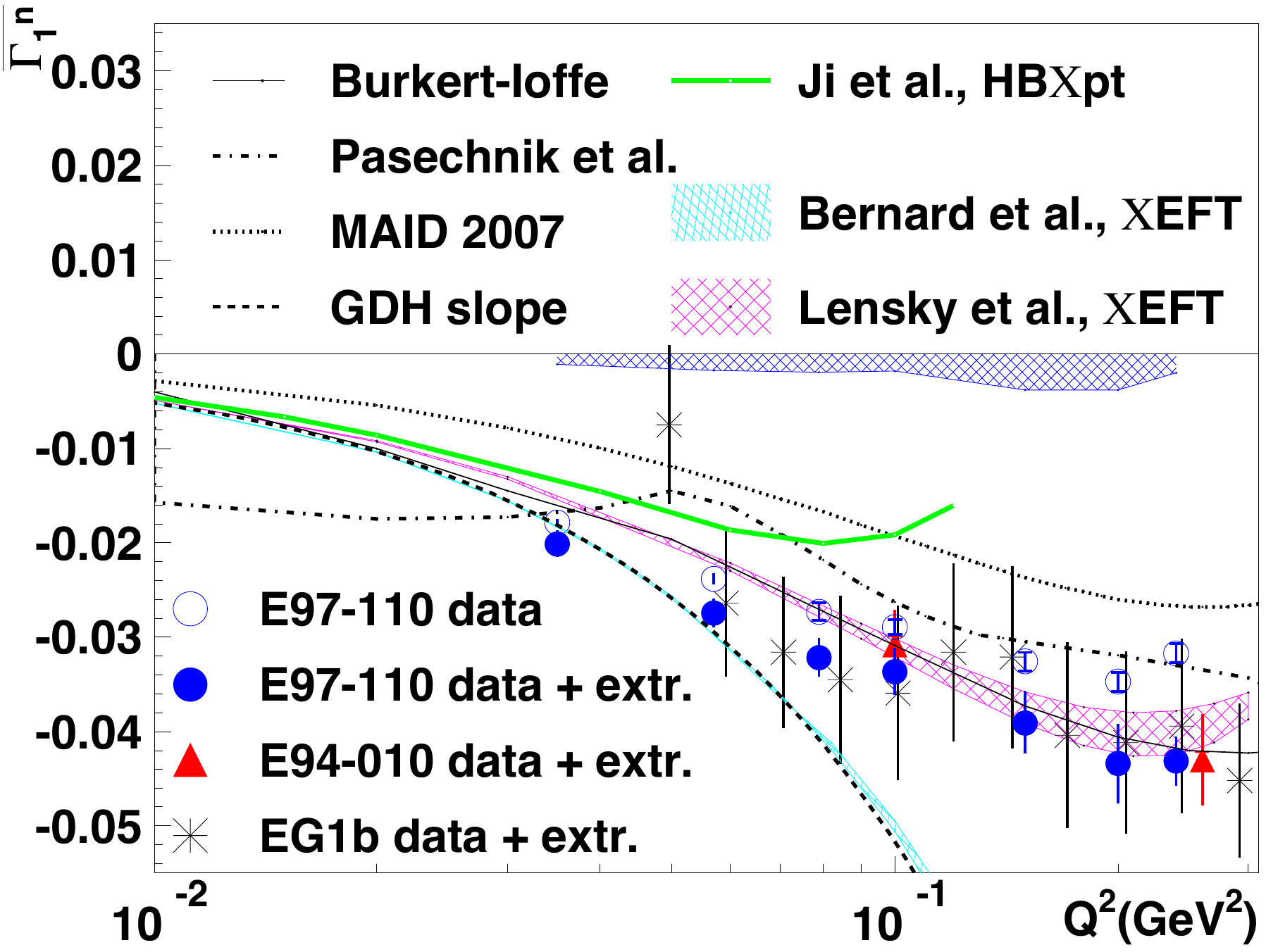}
 \caption{\label{moments}  $\overline{\Gamma}_1^n$ versus $Q^{2}$ from this experiment (E97-110),  
 compared to models and earlier JLab data from E94-010 and EG1b.
The open circles show the measured partial integral. 
The filled circles show the full integral with a low-$x$ contribution estimation.
The inner error bars on the E97-110 and E94-010 points,
often too small to be visible, represent the statistical uncertainties. 
The combined statistical  and uncorrelated systematic uncertainties are shown 
by the outer error bars. 
The correlated systematic uncertainty is indicated by the band  and typically is about half of the total
uncertainty.
The GDH sum rule provides  $d\overline{\Gamma}_1/dQ^2$ at $Q^2=0$ (dashed line), see Eqs.~\ref{eq:gdhsum_def1}
or~\ref{eq:gengdh}.}
\end{figure}

Two $\chi$EFT calculations have become available recently~\cite{Bernard:2012hb,Lensky:2014dda}, 
improving on the earlier ones~\cite{Bernard:2002pw,Ji:1999pd}.
Those had used different approaches, and different ways to treat for the $\Delta(1232)$ degree of freedom,
a critical component of $\chi$EFT calculations for baryons. For comparison, we also show in Fig.~\ref{moments} the older 
calculation~\cite{Ji:1999pd} in which the $\Delta(1232)$ is not accounted for.
The two state-of-art calculations~\cite{Bernard:2012hb, Lensky:2014dda} 
account explicitly for the $\Delta$ by computing the $\pi--\Delta$ graphs, but differ in their expansion methods for these corrections
and thus on how fast their calculations converge. Comparing them to our data will help to 
to validate the $\chi$EFT approach and determine the most efficient calculation technique.
Our $\overline{\Gamma}_1^n$ data agree with both calculations up to $Q^2\approx0.06$ GeV$^2$,
although a $\sim1.5\sigma$ offset exists between the calculation~\cite{Lensky:2014dda} and the data. They 
then agree only with 
calculation~\cite{Lensky:2014dda}, which predicts the plateauing of the data. 
The deviation for $Q^2 \gtrsim 0.06$ GeV$^2$ between data and the calculation from Ref.~\cite{Bernard:2012hb} is expected 
since, as pointed out in~\cite{Bernard:2012hb}, a similar deviation is seen with proton data but not for the isovector quantity 
$\Gamma_1^{(p-n)}$~\cite{Deur:2004ti}. 
The issue thus affects isoscalar combinations and can be traced to the later onset of loop contributions 
for isoscalar quantities (3 pions, in contrast with 2 pions threshold to isoscalar quantities)~\cite{Bernard:2012hb}.

\begin{figure}
  \centering
\includegraphics[width=0.6\textwidth]{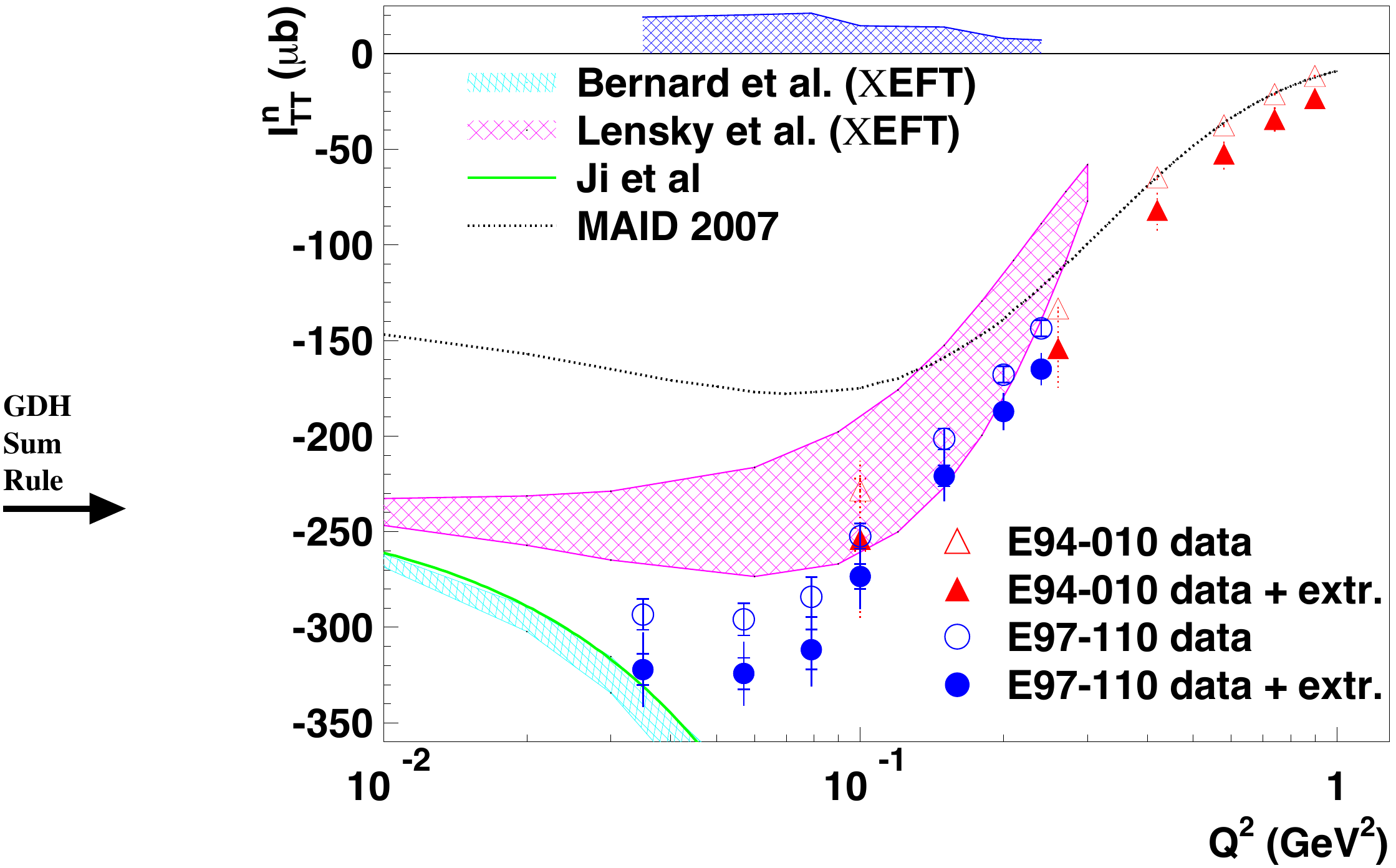}
\caption{\label{gdh_integral}  $I^n_\mathrm{TT}(Q^2)$ with 
  (filled circles) and without (open circles) the estimated unmeasured low-$x$ contribution.  
  The meaning of the inner and outer error bars and of the band is the same as in Fig.~\ref{moments}.
  Also shown are $\chi$EFT results, MAID (solid line) and earlier E94-010 data~\cite{Amarian:2002ar}.
    }
\end{figure}

$I^n_\mathrm{TT}(Q^2)$ is shown in Fig.~\ref{gdh_integral}.  The integration using only our data, and that with 
an estimate of the unmeasured low-$x$ part are represented by the open and solid circles, respectively.  
The open circles should be compared to the MAID result (solid line), which is larger
than the data. Our data and the earlier E94-010 data~\cite{Amarian:2002ar} are consistent.
As $Q^2$ decreases, our results drop to around $-325$~$\mu$b, agreeing with the 
$\chi$EFT calculation from Bernard~{\it et~al.}~\cite{Bernard:2012hb} and the earlier one from Ji~{\it et~al.}\cite{Ji:1999pd}. 
The calculation from Lensky~{\it et~al.}~\cite{Lensky:2014dda} displays the same $Q^2$-dependence as the data
but with a systematic shift.  
 Extrapolating the data to
$Q^2=0$ to check the original GDH sum rule is difficult since the calculations that could be used to guide the extrapolation markedly disagree. 
Data at lower $Q^2$ or a theoretical consensus on the $Q^2$-dependence of $I_{TT}^n$ are needed
to address the validity of the original GDH sum rule on the neutron.

\begin{figure}
  \centering
    \includegraphics[width=0.55\textwidth]{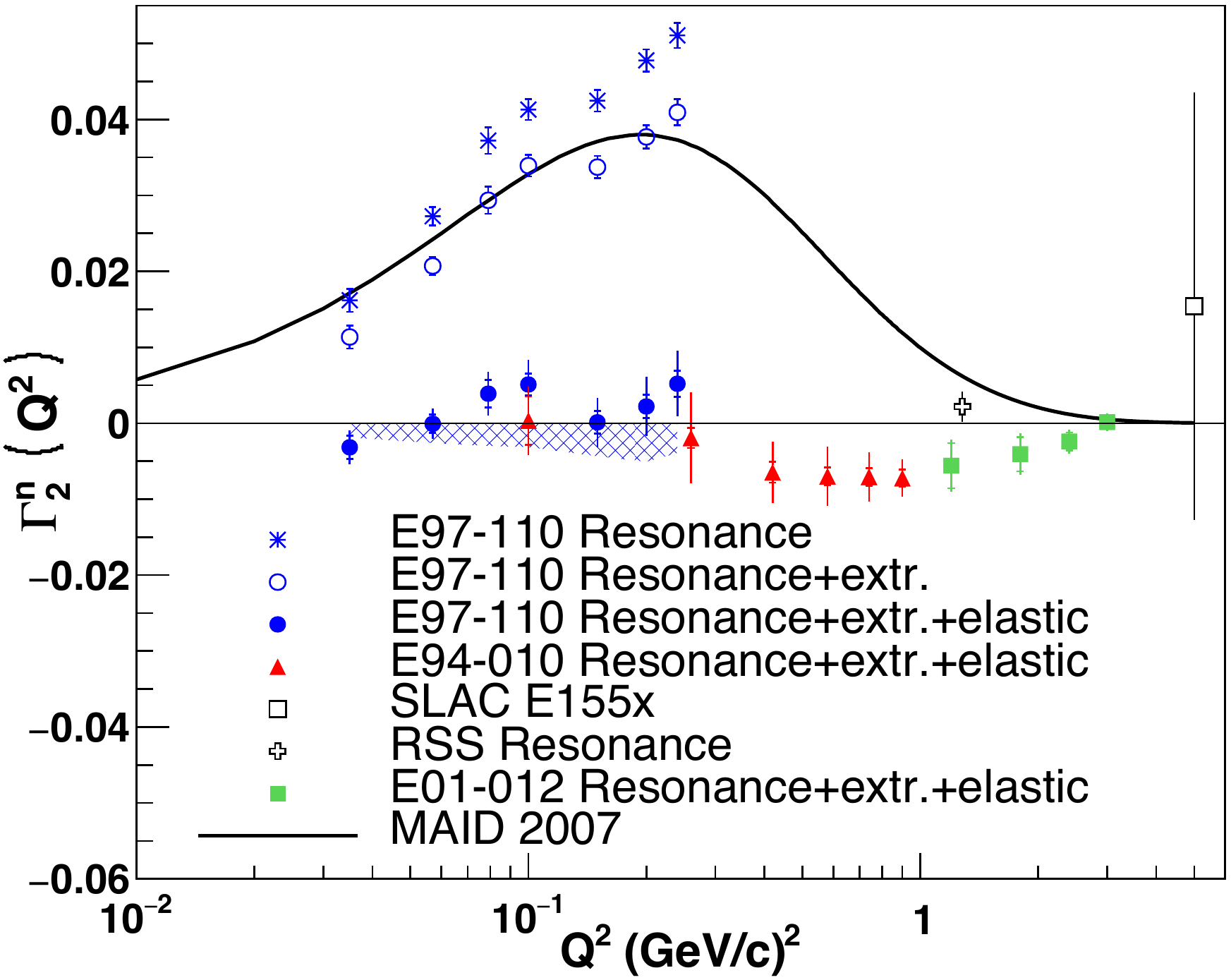} \\[\abovecaptionskip]
 \caption{\label{Gamma2}  $\Gamma_2^n$ versus $Q^{2}$. 
 The error band represents the correlated systematic uncertainty from radiative corrections, interpolation of 
$g_2$ to constant $Q^2$, model uncertainties in the neutron extraction from $^3$He, and the elastic contribution uncertainty.
The correlated systematic uncertainty typically represents about half of the total
uncertainty.
The uncorrelated systematic and statistical uncertainties added in quadrature are shown by the outer error bars. 
The inner error bars (when visible) represent the statistical uncertainty.
 Also shown is the MAID model with only resonance contributions.}
\end{figure}

$\Gamma_{2}^{n}\left(Q^{2}\right)$ is shown in Fig.~\ref{Gamma2}. 
The stars show the measured integral without low-$x$ extrapolation for the neutron, to be compared with  
MAID. This model underestimates the 
higher $Q^2$ data but agrees well at lower $Q^2$.  
The open circles represent the integral including an estimate for the low-$x$
contribution assuming $g_{2}$ = $g_{2}^{WW}$~\cite{Anthony:2002hy}, where $g_{2}^{WW}$ is the
twist-2 part of $g_{2}$~\cite{Wandzura:1977qf}.  
This procedure is used since there are little data to constrain $g_2$ at low-$x$. Since it is unknown
how well $g_{2}^{WW}$ matches $g_2$ there, one cannot reliably assess 
an uncertainty on the low-$x$ extrapolation and none was assigned.
The solid circles show the full integral with the elastic contribution evaluated using Ref.~\cite{Mergell:1995bf}.  
These data allow us to investigate the BC sum rule in this low-$Q^{2}$ region with the caveat of 
the unknown uncertainty attached to the low-$x$ extrapolation. Under this provision, 
the data are consistent with the sum rule expectation that $\Gamma_{2}=0$ for all
$Q^{2}$. They also agree with the earlier results from E94-010 (triangles)~\cite{Amarian:2002ar}.
Higher $Q^{2}$ data from E01-012 (filled squares)~\cite{Solvignon:2013yun}, RSS (open crosses)~\cite{Slifer:2008xu}, 
and E155x (open square)~\cite{Anthony:2002hy} are also consistent with zero.

In conclusion, $^3$He spin structure functions $g_1(\nu,Q^2)$, $g_2(\nu,Q^2)$ and the
spin-dependent partial cross-section $\sigma_\mathrm{TT}(\nu,Q^2)$ were measured at low $Q^2$. 
The moments $\overline{\Gamma_1}\left(Q^{2}\right)$,  $\Gamma_2\left(Q^{2}\right)$
and $I_{\mathrm{TT}}\hspace{-0.06cm}\left(Q^{2}\right)$ 
of the neutron are extracted at 0.035~$\le$~$Q^{2}$~$\le$~0.24~GeV$^{2}$.  They are 
compared to two next-to-leading-order $\chi$EFT calculations from two 
separate groups, Bernard~{\it et~al.}~\cite{Bernard:2012hb} and Lensky~{\it et~al.} calculation~\cite{Lensky:2014dda}.
The $\overline{\Gamma}^n_1(Q^{2})$ and $I^n_{\mathrm{TT}}\hspace{-0.06cm}\left(Q^{2}\right)$ 
integrals agree with published data at higher $Q^{2}$.
The data on $\overline{\Gamma_1^n}$ agree reasonably with both recent  $\chi$EFT calculations. 
The data on $I^n_{\mathrm{TT}}$ disagree with the calculation~\cite{Lensky:2014dda}
and that of~\cite{Bernard:2012hb} except at the lowest $Q^2$ point. 
That the results for two recent  $\chi$EFT methods differ, and that they describe with different degrees of success 
the data underlines the importance of 
the $\Delta$ degree of freedom for spin observables and the sensitivity of $\chi$EFT to the consequent $\pi$-$\Delta$ terms. 
The earlier E94-010 data had triggered improvement of the $\chi$EFT calculations. Now, the precise
E97-110 data, taken in the chiral domain, show that yet further sophistication of $\chi$EFT is needed before spin observables can be satisfactorily described.
Our determination of $\Gamma_{2}^n\hspace{-0.06cm}\left(Q^{2}\right)$ agrees with the 
BC sum rule in this low-$Q^2$ region, with the proviso that $g_2^{WW}$ is used to 
assess the unmeasured low-$x$ part of $\Gamma_{2}^n$.
Analysis of data down to $Q^2=0.02$~GeV$^{2}$ taken at a different time under different conditions, 
which requires a different analysis, is currently ongoing.
These data and results on $\sigma'_\mathrm{LT}$, 
the spin polarizabilities $\gamma^n_0$ and $\delta^n_\mathrm{LT}$, and moments for $^3$He 
will be reported in future publications. All these data, when 
combined with results~\cite{Adhikari:2017wox} obtained on deuteron 
and future proton data~\cite{EG4SHORT}
taken at low $Q^2$, will yield further extensive tests of calculations from $\chi$EFT,
the leading effective theory of strong interactions at low $Q^2$,  and eventually to QCD 
once the lattice QCD calculations of the Compton amplitudes involved in the sum rules  becomes available.

\noindent {\bf{Acknowledgments}}
We acknowledge the outstanding support of the Jefferson Lab Hall A technical staff  and the Physics and Accelerator 
Divisions that made this work possible. 
We thank A. Deltuva, J. Golak, F. Hagelstein, H. Krebs, V. Lensky, U.-G. Mei{\ss}ner, V. Pascalutsa, 
G. Salm\`e, S. Scopetta and M. Vanderhaeghen for useful discussions and for sharing their calculations. 
This material is based  upon work supported by the U.S. Department of Energy, Office of Science, 
Office of Nuclear Physics under contract DE-AC05-06OR23177, and by the NSF under grant PHY-0099557.
%
%
%
%

\noindent {\bf{References}}

\newpage

\newgeometry{left=2cm}

\begin{table*}
{\bf Supplemental material}\\
Data tables and kinematics of Jefferson Lab experiment E97-110.  
\begin{center}
{\footnotesize{
\begin{tabular}{|c|c|c|c|c|c|c|} \hline
$Q^2$        &$x_{min}$ ($W_{max}$)    & $I_\mathrm{TT}^{n,~data}\pm$(syst)    & $I_\mathrm{TT}^{n,x_{min}>x>0.001}\pm$(syst)  & $I_\mathrm{TT}^{n,x<0.001}\pm$(syst)  & $I_\mathrm{TT}^{n,~data+extr.}\pm$(syst) &  stat. \\ \hline 
0.035 GeV$^2$   & 0.0112 (2.00 GeV)    & $(-293.3 \pm 24.7) \mu$b& (-24.6$\pm$2.1)$\mu$b& (-4.3$\pm$1.0)$\mu$b& $(-322.1 \pm 25.0) \mu$b& 8.1 $\mu$b\\   
0.057 GeV$^2$   & 0.0181 (2.00 GeV)   & $(-295.9 \pm 23.8) \mu$b&  (-25.4$\pm$2.2)$\mu$b& (-3.1$\pm$0.8)$\mu$b&$(-324.3 \pm 24.0) \mu$b&8.4 $\mu$b\\ 
0.079 GeV$^2$   & 0.0249 (2.00 GeV)   & $(-284.2 \pm 25.7) \mu$b&  (-25.0$\pm$2.1)$\mu$b& (-2.5$\pm$0.6)$\mu$b&$(-311.7 \pm 25.9) \mu$b&10.5 $\mu$b\\ 
0.100 GeV$^2$   & 0.0183 (2.50 GeV)   & $(-252.4 \pm  20.9) \mu$b&  (-19.0$\pm$1.7 )$\mu$b& (-2.1$\pm$0.6)$\mu$b&$(-273.5 \pm  21.1) \mu$b& 6.5 $\mu$b\\ 
0.150 GeV$^2$   & 0.0273 (2.50 GeV)   & $(-201.6 \pm 17.9) \mu$b&   (-17.8$\pm$1.5 )$\mu$b& (-1.6$\pm$0.5)$\mu$b&$(-221.0 \pm 18.0) \mu$b& 5.5 $\mu$b\\ 	
0.200 GeV$^2$   & 0.0398 (2.40 GeV)  & $(-167.8 \pm 11.4)\mu$b&    (-18.1$\pm$1.5 )$\mu$b& (-1.3$\pm$0.4)$\mu$b& $(-187.3 \pm 11.6)\mu$b& 4.2 $\mu$b\\ 	
0.240 GeV$^2$   & 0.0547 (2.25 GeV)   &$(-143.7 \pm 9.8) \mu$b&  (-20.2$\pm$1.6  )$\mu$b& (-1.2$\pm$0.4)$\mu$b&$(-165.0 \pm 10.0)\mu$b& 4.2 $\mu$b\\ \hline 	
\end{tabular}
}}
\caption{Data and kinematics for $I_\mathrm{TT}^n(Q^2)$. From left to right: 
Four-momentum transfer; 
mimimum $x$ value experimentally covered (equivalent maximum $W$);
$I_\mathrm{TT}^n$ measured down to $x_{min}\pm$ systematic uncertainty;
estimated contribution to $I_\mathrm{TT}^n$ from $x_{min}$ to $x=0.001$ with systematic uncertainty;
estimated contribution to $I_\mathrm{TT}^n$ from $x<0.001$ with systematic uncertainty;
full $I_\mathrm{TT}^n$ with low-$x$ (equivalently large-$W$) extrapolation with systematic uncertainty;
statistical uncertainty on $I_\mathrm{TT}^n$.
The systematic uncertainties on the estimated low-$x$ contributions are added in quadrature to the experimental systematic uncertainty.
}\label{tab:results1}
\end{center}
\end{table*}
\begin{table*}
\begin{center}
{\footnotesize{
\begin{tabular}{|c|c|c|c|c|c|} \hline
$Q^2$   &$\Gamma_1^{n,~data}\pm$(syst)&$\Gamma_1^{n,x_{min}>x>0.001}\pm$(syst)&$\Gamma_1^{n,x<0.001}\pm$(syst)&$\Gamma_1^{n,~data+extr.}\pm$(syst)&  stat.      \\ \hline 
0.035 GeV$^2$      &$(-1.784 \pm 0.176)\times 10^{-2}$&$(-0.192\pm0.016)\times 10^{-2}$&$(-0.033\pm0.007)\times 10^{-2}$&$(-2.010 \pm 0.178)\times 10^{-2}$&$4.1\times 10^{-4}$   \\ 
0.057 GeV$^2$      &$(-2.379 \pm  0.232)\times 10^{-2}$&$(-0.323\pm0.027)\times 10^{-2}$&$(-0.039\pm0.010)\times 10^{-2}$& $(-2.742 \pm 0.236)\times 10^{-2}$&$5.8\times 10^{-4}$   \\ 
0.079 GeV$^2$     &$(-2.731 \pm 0.270)\times 10^{-2}$&$(-0.443\pm0.036)\times 10^{-2}$&$(-0.043\pm0.011)\times 10^{-2}$&$(-3.217  \pm 0.275)\times 10^{-2}$&$9.5\times 10^{-4}$  \\ 
0.100 GeV$^2$     &$(-2.892 \pm 0.303)\times 10^{-2}$&$(-0.424\pm0.038)\times 10^{-2}$&$(-0.047\pm0.013)\times 10^{-2}$&$(-3.364  \pm 0.308)\times 10^{-2}$&$8.0\times 10^{-4}$  \\ 
0.150 GeV$^2$     &$(-3.258 \pm 0.491)\times 10^{-2}$&$(-0.597\pm0.052)\times 10^{-2}$&$(-0.053\pm0.016)\times 10^{-2}$&$(-3.908  \pm 0.496)\times 10^{-2}$&$9.97\times 10^{-4}$ \\ 
0.200 GeV$^2$     &$(-3.473 \pm 0.554)\times 10^{-2}$&$(-0.808\pm0.065)\times 10^{-2}$&$(-0.058\pm0.018)\times 10^{-2}$&$(-4.339  \pm 0.561)\times 10^{-2}$&$10.0\times 10^{-4}$ \\  
0.240 GeV$^2$    &$(-3.169 \pm 0.301)\times 10^{-2}$&$(-1.084\pm0.083)\times 10^{-2}$&$(-0.062\pm0.020)\times 10^{-2}$&$(-4.315  \pm 0.320)\times 10^{-2}$&$10.2\times 10^{-4}$ \\ \hline 	
\end{tabular}
}}
\caption{Same as table~\ref{tab:results1} but for $\Gamma_1^n(Q^2)$. The $x_{min}$ (or $W_{max}$) values are the same as in table~\ref{tab:results1}.
}\label{tab:results2}
\end{center}
\end{table*}

\begin{table*}
\begin{center}
\begin{tabular}{|c|c|c|} \hline
$E$          &$E'$ range                 &$\theta$      \\ \hline 
1.147 GeV      &(0.445 -- 1.147) GeV            &9.03$^\circ$             \\             
2.135 GeV      &(0.905 -- 2.135) GeV            &6.10$^\circ$           \\              
2.234 GeV      &(0.937 -- 2.234) GeV            &9.03$^\circ$         \\           
2.845 GeV      &(0.955 -- 2.845) GeV            &6.10$^\circ$             \\ 
3.319 GeV      &(1.641 -- 3.217) GeV            &9.03$^\circ$             \\ 
3.775 GeV      &(0.827 -- 3.212) GeV            &9.03$^\circ$               \\ 
4.209  GeV     &(1.041 -- 3.288) GeV            &6.10$^\circ$              \\ 
4.404  GeV     &(1.420 -- 3.252) GeV            &9.03$^\circ$             \\ \hline   
\end{tabular}
\caption[E97-110 Kinematics]{Kinematics of the experiment. From left to right columns: 
beam energy  $E$;
range of transferred energy $E'$;
average scattering angle $\theta$;
}\label{tab:kines}
\end{center}
\end{table*}

\restoregeometry

\end{document}